\documentclass{easychair}

\usepackage{doc}
\usepackage{algorithm}
\usepackage[noend]{algpseudocode}
\usepackage{multirow}
\usepackage{subfigure}

\usepackage{listings}
\usepackage{color}
\usepackage{graphicx}

\definecolor{codegreen}{rgb}{0,0.6,0}
\definecolor{codegray}{rgb}{0.5,0.5,0.5}
\definecolor{codepurple}{rgb}{0.58,0,0.82}
\definecolor{backcolour}{rgb}{0.95,0.95,0.92}

\lstdefinestyle{mystyle}{
  backgroundcolor=\color{backcolour},   commentstyle=\color{codegreen},
  keywordstyle=\color{magenta},
  numberstyle=\tiny\color{codegray},
  stringstyle=\color{codepurple},
  basicstyle=\footnotesize,
  breakatwhitespace=false,
  breaklines=true,
  captionpos=b,
  keepspaces=true,
  numbers=left,
  numbersep=5pt,
  showspaces=false,
  showstringspaces=false,
  showtabs=false,
  tabsize=2
}

\title{Improving tasks throughput on accelerators using OpenCL command concurrency%
\thanks{This work has been supported by the Ministry of Education of Spain (TIN2013-42253P) and the Junta de Andaluc\'ia of Spain (TIC-1692)}}

\author{A.J.~L\'azaro-Mu\~noz\inst{1} \and
J.M.~Gonz\'alez-Linares\inst{1} \and
J. G\'omez-Luna\inst{2} \and
N. Guil\inst{1}
}

\institute{
 Dep. of Computer Architecture \\
    University of M\'alaga, Spain\\
    \email{nguil@uma.es}
\and
    Dep. of Computer Architecture and Electronics\\
    University of C\'ordoba, Spain \\
   }

\authorrunning{L\'azaro-Mu\~noz, Gonz\'alez-Linares, G\'omez-Luna and Guil}

\titlerunning{Improving task throughput in OpenCL}

\begin{document}

\maketitle

\begin{abstract}

  A heterogeneous architecture composed by a host and an accelerator must frequently deal with situations where several independent tasks are available to be offloaded onto the accelerator. These tasks can be generated by concurrent applications executing in the host or, in case the host is a node of a computer cluster, by applications running on other cluster nodes that are willing to offload tasks in the accelerator connected to the host. In this work we show that a runtime scheduler that selects the best execution order of a group of tasks on the accelerator can significantly reduce the total execution time of the tasks and, consequently, increase the accelerator use. Our solution is based on a temporal execution model that is able to predict with high accuracy the execution time of a set of concurrent tasks launched on the accelerator. The execution model has been validated in AMD, NVIDIA, and Xeon Phi devices using synthetic benchmarks. Moreover, employing the temporal execution model, a heuristic is proposed which is able to establish a near-optimal tasks execution ordering that signiﬁcantly reduces the total execution time, including data transfers. The heuristic has been evaluated with five different benchmarks composed of dominant kernel and dominant transfer real tasks. Experiments indicate the heuristic is able to find always an ordering with a better execution time than the average of every possible execution order and, most times, it achieves a near-optimal ordering (very close to the execution time of the best execution order) with a negligible overhead. Concretely, our heuristic obtains, on average for all the devices, between 84\% and 96\% of the improvement achieved by the best execution order.

\end{abstract}



%
%

\section{Introduction}
\label{sect:introduction}

Current heterogeneous platforms include latency-oriented CPUs and throughput-oriented accelerators that are specialized on different types of workloads. Typically, massively parallel computations are better suited for accelerators, while sequential or moderately parallel ones run faster on CPUs. This way, different parts of real applications might run on different types of processors. This entails a need for data movement between memory spaces over the PCIe bus, which poses an inherent penalization on performance. However, it is possible to alleviate this bottleneck when several independent tasks (also called kernels) are ready to be launched, since data transfer and kernel computation commands from different tasks can be overlapped.

Some Application Programming Interfaces (API) such as CUDA~\cite{CUDA} and OpenCL~\cite{OPENCL} provide features to overlap communication between CPU (namely the host) and accelerator (the device) with computation, by employing CUDA streams or OpenCL command queues (CQ), respectively. Overlapping commands increases both tasks throughput\footnote{Tasks throughput is defined as the number of tasks executed on the accelerator per time unit.} and accelerator use by reducing idle periods between tasks executions.

\begin{figure}[ht]
  \centering
  \includegraphics[width=1.0\linewidth, height=0.4\linewidth]{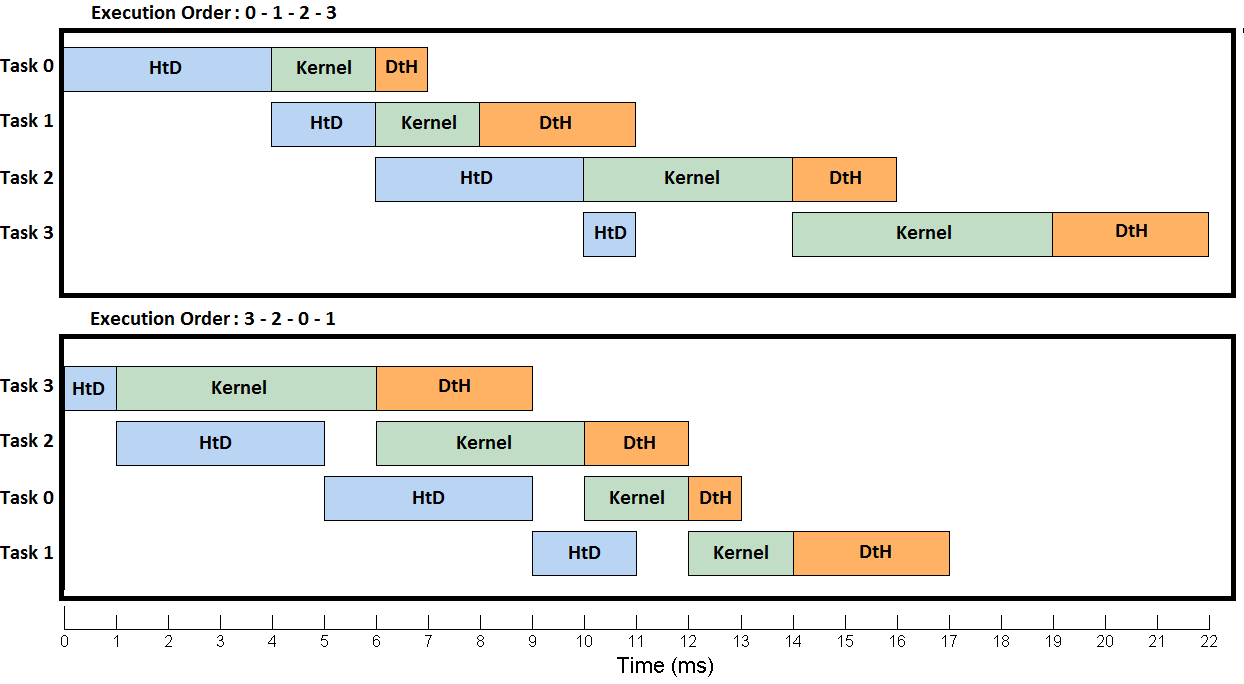}

\caption{Two examples for the concurrent execution of the same four tasks with two different orderings on a generic device. The host to device transfers, kernels computation and device to host transfers are represented by blue, green and orange boxes respectively. Top and bottom views correspond to two different orderings in command queues execution.}
\label{fig:Examples}
\end{figure}

Concurrent applications running on a
host can simultaneously submit several kernels to the accelerator. In a similar way, current GPU clusters employ frameworks, like rCUDA or CUDA MPS \cite{rCUDA,CUDA_MPS}, that allow
processes to execute kernels in
a remote accelerator. Consequently, many simultaneous tasks can be found in the host, ready for offloading onto the accelerator. The achieved throughput of a group of tasks ($TG$) ready to be submitted to the accelerator depends on the order the tasks are
launched as the scheduling policy affects the final overlapping degree. This fact is illustrated in Figure~\ref{fig:Examples}, where the execution timeline of four offloaded tasks employing two different orders is shown. Note that, in this example, transfers from host to device ($HtD$) and device to host ($DtH$) can also overlap.

The previous example shows that the order in which tasks are submitted to a device might have an important impact on the total execution time. Assuming several tasks are simultaneously available for offloading, finding the optimal order would require to test all possible orderings and choose the one resulting in the shortest execution time. However, this brute-force approach is not feasible in runtime since testing all possible combinations for $N$ independent tasks involves evaluating $N!$ different orderings. In this paper this issue is addressed by proposing a runtime approach that selects a near-optimal ordering for concurrent task offloading using OpenCL CQs. The main contributions of our proposal are the following:
\begin{itemize}
 \item Given a specific ordering of an arbitrary set of tasks, an event-driven simulator is proposed which is able to predict with high accuracy the total execution time of those tasks on a device. Our model takes into account that kernel commands can overlap transfer commands, and data transfers can simultaneously occur in opposite directions (HtD and DtH).
 \item Based on the previous simulator a runtime system is proposed, consisting of both a host proxy thread that receives tasks to be offloaded and a heuristic that is able to calculate a near-optimal ordering of these tasks.
\end{itemize}

Exhaustive experimentation has been conducted employing both synthetic and real tasks benchmarks on Intel, NVIDIA, and AMD devices that demonstrates the accuracy and applicability of our proposal.

The rest of the paper is organized as follows. Next section discusses the most significant related works. Section~\ref{sec:openclCQ} explains how OpenCL CQs can be used to overlap commands on devices with one and two DMA engines. Section~\ref{sec:simulator} describes the simulator we have developed, that predicts the total execution time of a group of concurrent tasks on a device given a specific ordering. Section~\ref{sec:ExecutionOrder} presents the proposed runtime heuristic that is able to find a near-optimal. Section~\ref{sec:results} conducts exhaustive experiments employing several synthetic and real benchmarks containing both execution and transfer bound tasks, and demonstrates the benefits of the proposed heuristic. Finally, Section~\ref{sec:conclusion} states the conclusions of our work.

\section{Related works}
\label{sec:related}
Current accelerators APIs, such as CUDA and OpenCL, provide a set of software queues where host threads can submit transfer and execution commands to run a task on an accelerator. These software queues are named Streams in CUDA~\cite{CUDA} and Command Queues~\cite{OPENCL} in OpenCL. Thanks to this software queues, concurrency can be increased on accelerators. Thus, {\it temporal sharing} of resources, where transfer and execution commands from different tasks overlap, can take place. In addition, {\it spatial sharing} can be also feasible because several tasks could be simultaneously executed in an accelerator by partitioning the available computing units. Hardware support to enhance both types of resources sharing is available in different devices, as HyperQ~\cite{CUDA} (NVIDIA) and ACEs~\cite{AMD} (AMD). As our approach can take advantage of both types of resource sharing, next we discuss the most relevant works in these topics.

\subsection{Temporal sharing}

As data transfers can have an important impact in the total execution time of a task on a accelerator, there have been several efforts in modeling data transfers and kernels computation on a GPU to improve overall performance. A GPU performance modeling framework that takes into account both kernel execution time and data transfer time is discussed in~\cite{Boyer2013}. In that work, the amount of data to be transferred for a sequence of kernels is predicted and a simple performance model of the PCIe bus is used to determine how long the data transfer takes. In~\cite{Werkhoven2014} a more precise model for CPU-GPU data transfers using PCIe is proposed. They identify three NVIDIA GPU categories: (1) devices with implicit synchronization and 1 copy engine (Fermi architecture GPUs and some Kepler cards as the GTX 680), (2) devices with no implicit synchronization and 1 copy engine (GTX Titan), and (3) devices without implicit synchronization and 2 copy engines (Tesla K20c, S2050). They obtain an approximate performance model for each category that includes what extent of overlap can be achieved between transfers and kernels computation.

Other works have studied how to improve the performance of streamed execution. In~\cite{Gomez-Luna2012} streamed executions are modeled to obtain the optimal number of streams once kernel execution and data transfers times are known, and this optimal number can be recomputed on-the-fly according to the current workload of the kernel. In~\cite{Liu2015} it is discussed how data partition and scheduling can influence the achieved performance for pipelining schedules in NVIDIA and AMD devices. However in their study the overlapping between transfers in different directions, happening in modern devices with two DMA copy engines, is not taken into consideration. Recently, Li et al.~\cite{Zhaokui:2016} use hStreams \cite{Souza:2015} to study the performance obtained on heterogeneous MIC-based platforms. They carry out a qualitative analysis of the execution time improvement achieved when temporal and spatial overlapping is employed.

\subsection{Spatial sharing}

Concurrent kernel execution (CKE) is a GPU feature that allows the execution of several concurrent kernels on a single device. The execution of concurrent kernels is subjected to the availability of GPU resources, namely number of registers, maximum number of threads, shared memory capability and maximum number of resident blocks. Early works in this topic tried to reorganize kernel launching to take advantage of CKE. Thus, Wang et al. \cite{Wang112} proposed a context funneling scheme instead of the typical context switching approach to execute lightweight kernels on GPUs. This technique allowed multithreaded applications to efficiently share a CUDA context in CUDA versions prior to 4.0 and take advantage of CKE feature. Later,  Wende et al. \cite{Wende2012} proposed a producer-consumer mechanism to reorganize the kernel launching of concurrent threads on devices with only one hardware queue for kernel management (NVIDIA Fermi architecture). In this scheme the consumer associates each CPU thread (producer) to a different CUDA stream. Then, different streams are consecutively located in the hardware queue to increase the concurrence.  In \cite{Awatramani13}, Awatramani et al. employ GPGPUSim to illustrate how the concurrent execution of memory and compute bound kernels on the same Stream Multiprocessor (SM) can help to increase the utilization of computational units and reduce the load of the memory subsystem.

Since the launch of NVIDIA Kepler and AMD Southern Islands architectures, streams (or command queues) can  be directly mapped into several hardware managed queues removing false dependencies between kernels that could appear when only a hardware queue was available. Based on this new feature, several authors have presented novel kernel scheduling techniques to take advantage of CKE. Thus, Zhong et al. \cite{Zhong:2014:KHG:2707571.2707642} propose a scheme to slice heavyweight kernels into a set of lightweight kernels that could be concurrently executed. They also develop a scheduling mechanism that selects, by pairs, kernels to be co-scheduled. Suzuki et al.~\cite{Suzuki15} presented a runtime library that checks the GPU memory requirements of concurrent kernels and carries out application suspension by swapping them out to CPU memory. These papers are only focused on exploring mechanisms to overlap kernels execution and they do no study the impact that data transfer commands can have in the execution of applications.

In this work we present an efficient solution to the execution of concurrent tasks on GPUs by improving the current models that overlap data transfers and execution commands. This way, we can tackle more complex and realistic scenarios, like multithreaded scenarios where several CPU threads can launch several concurrent tasks with varying time requirements.

\section{Asynchronous command execution in OpenCL}
\label{sec:openclCQ}
The execution of a specific kernel in a heterogeneous platform requires offloading a task from the host onto the device. This offloading implies the execution of a kernel (\emph{K}). In addition, it can also include host-to-device transfers (\emph{HtD}) of input data required by the kernel, and device-to-host transfers (\emph{DtH}) of output data produced by the kernel execution. In this section we explain how OpenCL asynchronous commands for data transfers and kernel computation must be employed to take advantage of the device hardware. The explanation includes memory allocation policies of data to be transferred, and  management of command queues (CQs).

\subsection{Memory allocation in OpenCL}

 Data transfers between CPU and accelerator require the allocation of CPU memory. Different vendors programming guidelines recommend to use page-locked memory for exploiting the maximum interconnect bandwidth~\cite{CUDA}. These memory pages remain in main memory and are never paged out. According to Margiolas et al.~\cite{Margiolas:2014} the allocation of these buffers can improve communication performance but it may also yield longer allocation times. In order to obtain both good transfer and allocation times they propose different allocation policies to be tested on the specific platform. In the experiment we have conducted the \emph{OpenCL} policy obtains the best result. It  uses the OpenCL runtime to allocate locked memory segments on the host. Then, these memory segments are attached to the application address space.

\subsection{Command Queue management}

 A CQ is a software queue used by the host application to submit commands to the device. Usually, these commands are requests to initiate memory transfers or execute kernels. Commands submitted to the same CQ are executed in order. Thus, command overlapping can take place when the host application sends asynchronous commands to different CQs. The number and type of commands that can be simultaneously executed depend on the device computer capabilities. Thus, if the device has a single DMA engine, \emph{HtD} and \emph{DtH} commands could not be overlapped even though both commands were placed in different CQs. Similarly, the execution of several kernels could not be overlapped if one of the kernel exhausts the device computation resources, despite the fact that the device could support concurrent kernel execution (CKE).

 CQs also allow to associate events to commands. The host thread can check the status of the event to know if a command has been completed. This way a host thread can establish dependencies between commands executing in different CQs by launching one command upon termination of a previous one.

 The way commands of a specific task must be submitted to CQs to take advantage of concurrent execution depends on the mapping between CQs and hardware queues on the device. Thus, CUDA programmers typically send the task commands to the same CQ, which is named \emph{Stream} in CUDA terminology, relying on Hyper-Q capabilities to overlap commands from different tasks and eliminate false dependencies among tasks. However, we have followed a different approach in OpenCL. It takes into account both the specific characteristics of the devices utilized in the experiments, and several restrictions that improve tasks throughput. Following we present two schemes to map commands into CQs that are based on the number of DMA engines employed by the device.

\begin{figure}[ht]
  \centering
  \includegraphics[width=1.0\linewidth, height=0.2\linewidth]{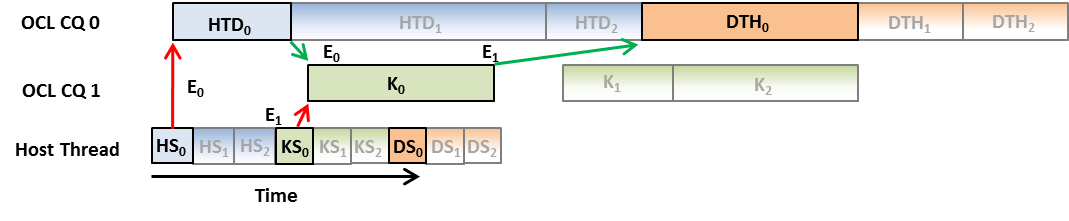}
\caption{Launching scheme for devices with one DMA engine. This scheme uses two OpenCL (OCL) command queues to asynchronously launch three tasks. One OCL command queue is used for memory transfers and another for computation. The host thread launches the commands grouping them by type. In addition, intra-task dependencies have to be managed by the host thread through OpenCL events.}
\label{fig:twoCQLaunching}
\end{figure}

 In Figure \ref{fig:twoCQLaunching} we depict the proposed scheme for launching three tasks in devices with one DMA engine as Intel Xeon Phi 5100 series. Our approach only uses two CQs and the host thread submits the commands grouping them by type (first all \emph{HtD} commands, then \emph{K} commands, and finally \emph{DtH} commands). Thus, all the transfer commands are sent to one queue (OpenCL CQ 0 in Figure \ref{fig:twoCQLaunching}). In addition the \emph{HtD} commands of all tasks are sent before the \emph{DtH} commands to reduce idle times in the DMA engine caused by dependencies between \emph{K} commands and \emph{DTH} commands. Kernel commands are submitted to the other OpenCL CQ (CQ1 in Figure~\ref{fig:twoCQLaunching}) to improve their overlapping with transfer commands. Note that more than one CQ could be employed to submit kernel commands and, this way, to grant CKE if possible.

 Since the memory transfer and kernel commands belonging to a task are launched on different CQs, our scheme must insert intra-task dependencies through OpenCL events. Figure~\ref{fig:twoCQLaunching} shows these dependencies. For the sake of clarity, they are only shown for task 0. Thus, when a \emph{HtD} or \emph{K} command is submitted by the host thread (indicated as HS$_{0}$ and KS$_{0}$), an OpenCL event is also associated to this command (indicated as E$_0$ and E$_1$ in Figure~\ref{fig:twoCQLaunching}). This event changes from the submitted state to the completed state when its execution is finished. Red and green arrows are drawn in Figure~\ref{fig:twoCQLaunching} to indicate the moment an event is submitted and completed, respectively. Hence, task 0 \emph{K} command execution is delayed until E$_{0}$ reaches the completed state. Similarly, task 0 \emph{DtH} command execution does not start until E$_{1}$ is completed.

\begin{figure}[ht]
  \centering
  \includegraphics[width=1.0\linewidth, height=0.2\linewidth]{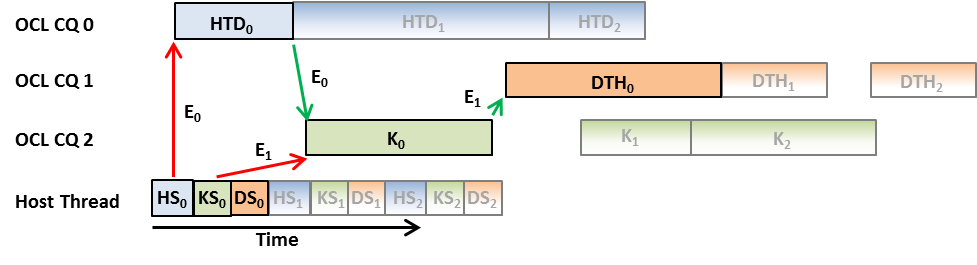}
  \caption{Launching scheme for devices with two DMA engines. This scheme uses three CQs to asynchronously launch three tasks. Two CQs are employed for memory transfers and the other one for computation. OpenCL runtime associates even and odd CQs to different DMA engines, so \emph{HtD} and \emph{DtH} commands are respectively launched on CQs 0 and 1. The host thread launches the commands grouping them by task. In addition, the intra-task dependencies are performed by the host thread using OpenCL events.}
  \label{fig:threeCQLaunching}
\end{figure}

For devices with two DMA engines such as AMD R9, the proposed scheme is shown in Figure~\ref{fig:threeCQLaunching}. This scheme uses three OpenCL CQs. Two queues are now employed for $HtD$ and $DtH$ commands because transfers in opposite directions can execute independently. The command queue associated with each data transfer is important as OpenCL runtime associates even and odd CQs command queues to different DMA engines~\cite{AMD}. Therefore, the \emph{HtD} and \emph{DtH} commands should be launched respectively on OpenCL CQ0 and CQ1 as shown in~\ref{fig:threeCQLaunching}. In this example one CQ is employed to submit kernel execution commands but CKE could be feasible by using additional CQs per kernel commands. Note that in this scheme the host thread submits commands in task order (all the commands of a task in a row). This allows to increase the amount of time in which both DMA engines are working simultaneously. The host thread is also responsible for handling  intra-task dependencies by inserting OpenCL events. Similarly to devices with one DMA engine, intra-tasks dependencies are inserted and managed so that a task command can start only when previous task commands have completed.

\begin{figure}[ht]
\centering
\subfigure[Model for devices with one DMA engine.]{\includegraphics[width=90mm]{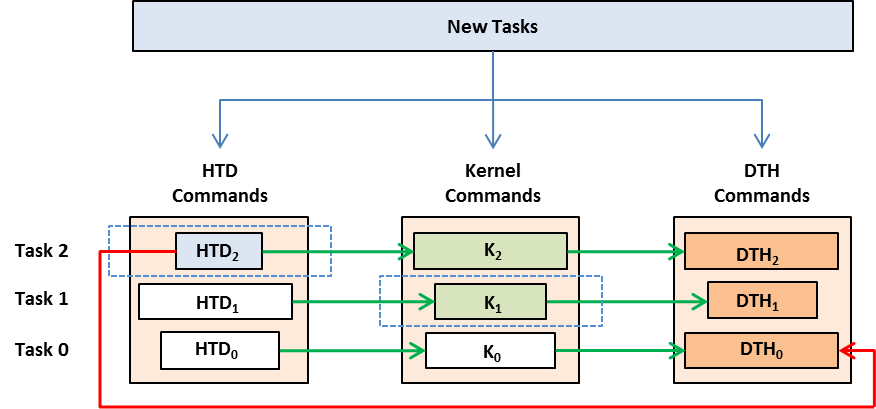}\label{fig:1copyEngineSim}}
\subfigure[Model for devices with two DMA engines.]{\includegraphics[width=90mm]{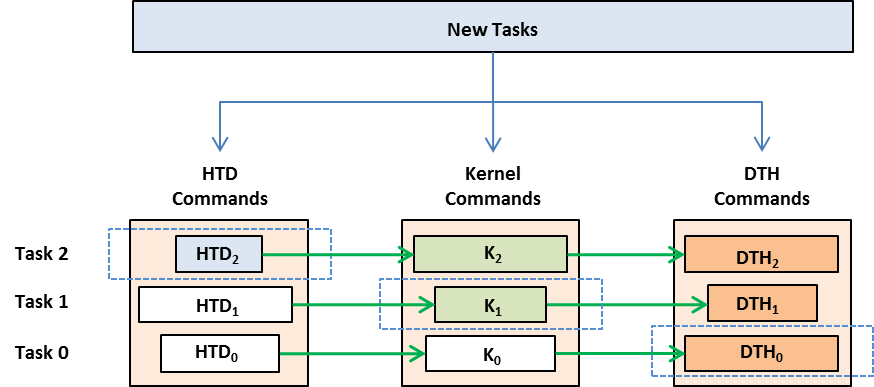}\label{fig:2copyEngineSim}}
\caption{Model execution is simulated using FIFO software queues. Green arrows between commands in different queues represent dependencies due to the order inside each task. Red arrows simulate the dependency when submitting $HtD$ and $DtH$ commands in a device with one DMA engine ($DtH$ commands are submitted just after all the $HtD$ commands have been sent). Commands are inserted in their respective queues, and launched in that order. Current commands in execution are inside dotted rectangles, while commands in white boxes have already been executed.} \label{fig:Sim}
\end{figure}

\section{Modelling the execution time of a group of independent tasks}
\label{sec:simulator}
Several threads or local processes running on a host can submit concurrent tasks to a specific accelerator connected through an I/O bus. Some frameworks, like rCUDA or CUDA MPS \cite{rCUDA,CUDA_MPS}, also allow remote processes to execute kernels in an accelerator. Consequently, many independent tasks can be simultaneously ready on the host node for offloading onto the accelerator. In this section we develop a model that can predict the total execution time of a group of independent tasks ($TG$) in a device, given a specific tasks execution order. Then, a heuristic is built on this model which is able to predict a near-optimal order for the $TG$ execution, effectively increasing tasks throughput.

\subsection{A generic model for task concurrent computation}
\label{sec:model}

As indicated in the previous section, a generic task is a sequence of three stages \emph{HtD-K-DtH} which must be executed in order. Each transfer stage can be null (i.e., no commands are executed) or composed by one or more commands.
%
Taking into account the OpenCL tasks submission schemes already explained, we propose a model with three FIFO software queues to simulate the computation of a $TG$. Each queue is devoted to the simulation of a different command type. Since there exist dependencies among the commands belonging to a task, these software queues are not independent. Figures~\ref{fig:1copyEngineSim} and \ref{fig:2copyEngineSim} depict the dependencies involved between commands belonging to different queues for devices with one and two DMA engines, respectively.
The head of each queue has been surrounded by a blue dotted rectangle. Thus, the simulation status shown in Figure \ref{fig:1copyEngineSim} indicates that $HtD_2$ and  $K_1$ commands are being executed (or ready to be executed). Similarly, white boxes make reference to commands that have already been executed, while the remaining commands are waiting to fulfil implicit (green arrow) and explicit dependencies (red arrow). This way, Figure \ref{fig:1copyEngineSim} shows that $DtH_0$ is not ready for execution because $HtD_2$ has not finished yet. This dependency simulates the behaviour of the proposed OpenCL scheme for submitting tasks in devices with one DMA engine. No explicit dependencies are present in Figure~\ref{fig:2copyEngineSim} as a device with two DMA engines is simulated. Moreover, $DtH_0$ command is being executed or ready to be executed.


\begin{figure}[ht]
  \centering
  \includegraphics[width=0.9\linewidth, height=0.5\linewidth]{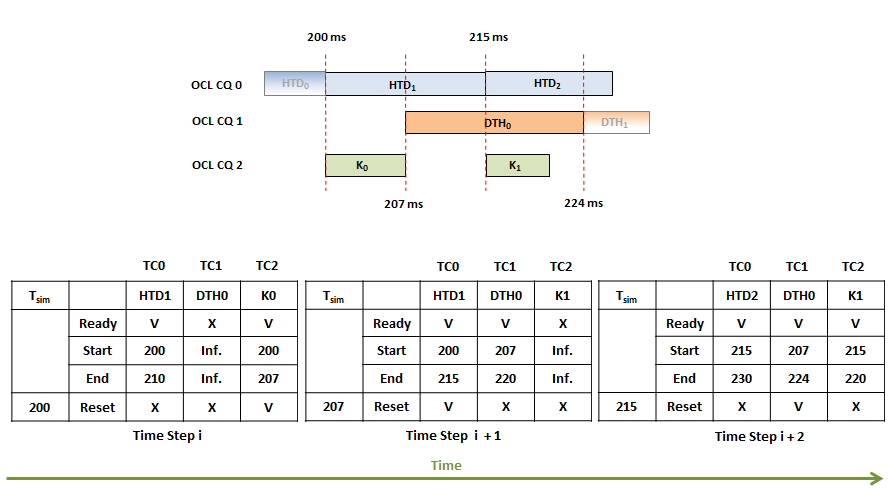}
  \caption{Example of the simulation timeline of several commands belonging to three different tasks. Vertical red lines identify the simulation steps. The information calculated in three of these simulation steps is also shown in the depicted tables. Thus, at each time step, the ready commands at the head of the FIFO queues are identified and their start and end execution times are calculated.}
  \label{fig:flowDiagram}
\end{figure}

Figure~\ref{fig:flowDiagram} shows an example explaining how the simulator performs the calculation of the total execution time of a $TG$. The device used in the figure has two DMA engines, thus three CQs are employed for commands submission. Three time counter structures are created ($TC0$, $TC1$, and $TC2$), one per each command type ($HtD$, $DtH$, and $K$), where the start and end times of the corresponding commands are annotated at each simulation step. The example starts at simulation time $T_{sim}$=200 ms (Time Step i in Figure~\ref{fig:flowDiagram}), that is, $HtD_0$ has just finished and, consequently, it has been removed from the head of the FIFO queue associated to $HtD$ transfers ($FIFO_{HtD}$). At this time two commands become ready in the head of their FIFO queues: $HtD_1$ that reaches the head of $FIFO_{HtD}$, and $K_0$, located at the head of $FIFO_K$, that becomes ready after fulfilling its dependence condition w.r.t. $HtD_0$. Then, the start and end times of the ready commands are estimated and annotated in $TC0$ and $TC2$ time counter structures. The estimation of the execution time of both commands is based on a previously established execution model that is explained in the following subsection. After that, simulation step moves forward to the earliest end time of the ready commands. Thus, in this example, the simulation time is updated with $T_{sim}$=207 ms, just when $K_0$ command finishes (Time Step i+1 in Figure \ref{fig:flowDiagram}). Now $K_0$ is removed from $FIFO_K$ and $DtH_0$ becomes ready as its dependency condition with $K_0$ is fulfilled. Next, the start and end times of the new ready command are annotated. However, a transfer overlap between $HtD_1$ and $DtH_0$ is detected, thus the overlapping degree is calculated and new end times for both commands are re-calculated and annotated in $TC0$ and $TC1$ (see next subsection for details). Note that after re-evaluation, the end time of $HtD_1$ has changed from 210 to 215. Once the end times have been correctly calculated, the simulator advances again to the earliest end time of the ready commands, that is, $T_{sim}$=215 (Time Step i+2 in Figure \ref{fig:flowDiagram}). At this simulation time, $HtD_1$ is de-queued from $FIFO_{HtD}$ and both $HtD_2$ and $K_1$ become ready. The simulator realizes that a new overlapping between $DtH_0$ and $HtD_2$ happens and, consequently, a re-estimation of the end time of both commands is performed. As a consequence, the end time of $DTH_0$ changes from 220 to 224.

As it can be observed in previous examples, our model avoids concurrent kernel execution, that is, only a CQ is employed for kernel execution commands. The reason for this is twofold.  On one hand the improvement obtained by CKE when kernels exhaust one or several device resources (register, shared memory, number of workgroups per multiprocessor, etc.) is very limited and these types of kernels are the most frequently executed on accelerators. In this situation, HyperQ (NVIDIA) only can overlap the tail of a kernel (just when some resources are being freed) with the start of another concurrent kernel. The improvement obtained by this small overlapping does not justify the increasing complexity of the model when more CQs are required (one additional CQ per concurrent kernel).  On the other hand, the mechanism employed by the devices to perform CKE has not been disclosed. This makes very difficult to develop a general model for concurrent kernel execution, which is required by our approach to estimate kernel commands execution time. Even so, as we show in the experimental results section, the predicted ordering derived from our model is able to beat most of the possible execution orders when CKE is set.



\subsection{Estimation of the command execution time}

The simulation performed by our model is based on a previous calculation of the execution time of each OpenCL command. Following we discuss how to accomplish this calculation.

\subsubsection{Transfer commands.}
\label{sec:transfer_model}

\begin{figure}[t]
  \centering
  \includegraphics[width=0.8\linewidth]{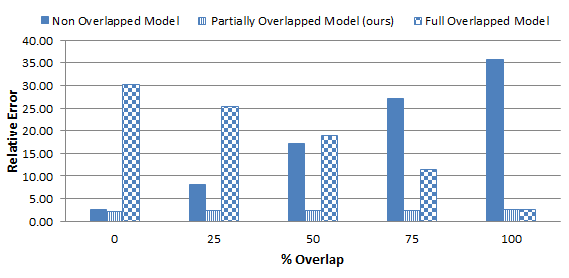}
  \caption{Relative error of the prediction for bidirectional transfers with varying degrees of overlapping in a AMD R9 card. Three prediction models are considered: non-overlapped transfers, partially overlapped transfers (ours) and full overlapped transfers.}
  \label{fig:overlap_error}
\end{figure}

The \emph{HtD} and \emph{DtH} transfers can be estimated with a PCI Express model like the one presented by Werkhoven et al.~\cite{Werkhoven2014}.  In this work the authors propose an extension of a PCIe transfer time model called LogGP~\cite{Culler1996, Alexandrov1997}, whose parameters can be measured by running a simple benchmark application. Estimation of two transfers in opposite directions is also considered by this model. The model works well when there is no overlap o there is a full overlap between two transfers commands, but it fails to simulate a partial overlap.  Thus, we have developed a more accurate model at any overlapping degree. Figure~\ref{fig:overlap_error} compares the relative error of the predicted time of our method, named partially overlapped model, in a AMD R9 device with both non-overlapped and full overlapped models. In this experiment, asynchronous transfers and pinned host memory are used. One CQ executes a {\em HtD} command while the other CQ launches a {\em DtH} command that overlaps 0\%, 25\%, 50\%, 75\% and 100\% with the other command. The experiment has been conducted using different transfer sizes, between 16 MB and 512 MB, to assess the accuracy of the prediction models. Then, the execution time of both transfers is measured and the error prediction is calculated for the three different models. The figure shows that our proposal obtains a relative error below 2\% and works better at any overlapping degree.

\subsubsection{Kernel commands.}
\label{sec:kernel_model}

There have been many efforts in the literature to characterize the performance of kernel commands. Some models are aimed to analyze the potential performance improvements of some optimization techniques ~\cite{Meng2011,Sim2012}. As improving kernel code is not the aim of this work, but to predict their execution time, we have employed a simple linear model as the one used by Liu et al.~\cite{Liu2015}. In that work the time of computing some $m$-size input data by a given kernel is modeled by

\begin{equation}
T = \eta \cdot m + \gamma
\label{eq:kernel_model}
\end{equation}

where $\eta$ is the computing rate defined by the computation time per unit data size, and $\gamma$ is the kernel invoking latency. Our concurrent execution tasks model only needs to keep a record of these two parameters based on an offline previous execution for each kernel we want to schedule. It is also possible to  profile every kernel and extract their execution time or use computation times gathered from previous executions as done by OmpSs and StartPU \cite{OmpSs,StartPU}.

\subsection{Model Validation}
\label{sec:model_val}

Our model has been be tested in three accelerators:  AMD R9, NVIDIA K20c and Intel Xeon Phi 5100 series. The main characteristics of these devices are shown in Table \ref{tb:platforms}. The evaluation of our model was carried out using a set of synthetic tasks with different transfers and computation time. Listing \ref{lst:SyntheticKernel} shows the simple code (a scalar-vector multiplication) of the kernel executed by these synthetic tasks. Parameter \emph{input} contains the pointer to the data array, whose size establishes the duration of the \emph{HtD} and \emph{DtH} commands. Similarly, different values for \emph{num\_iterations} lead to different kernel execution times. Table~\ref{tbl:synthTasks} describes the synthetic tasks designed to test our model. The values in this table are a percentage of a time unit, which has been adjusted to be 10 ms. For example, first task, $T0$, takes 1 ms for {\em HtD} stage, 8 ms for {\em K} stage and a 1 ms for {\em DtH} stage.

\begin{figure}[t]
  \centering
  \includegraphics[width=0.8\linewidth]{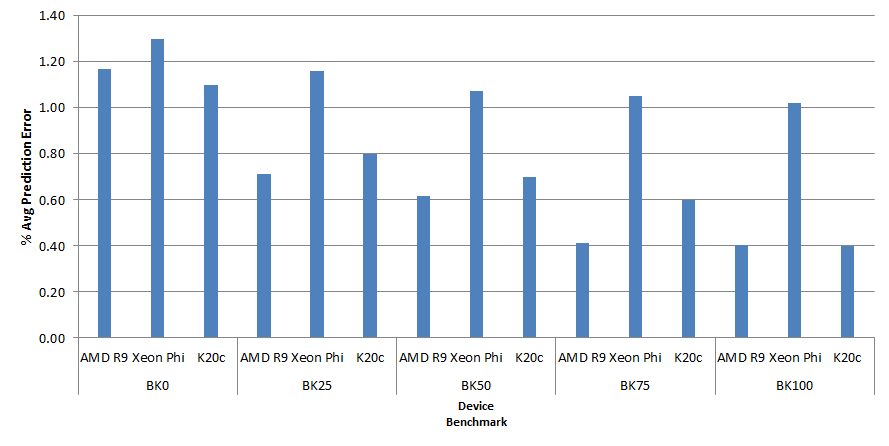}
  \caption{Average prediction error for all tasks permutations in each benchmark.}
  \label{fig:prediction_error}
\end{figure}

The synthetic tasks can be classified in two categories: 1) Dominant transfer tasks: when $t_{HtD}+t_{DtH}>t_K$ (DT tasks, $T4-T7$), and 2) Dominant kernel tasks: when $t_{HtD}+t_{DtH} \leq t_K$ (DK tasks, $T0-T3$). Considering this tasks classification, five benchmarks have been considered to test the model with the aim to have a high diversity workload. These benchmarks are shown in Table \ref{tbl:synthBenchmarks}. Each benchmark contains four different tasks whose percentage of DK tasks is included in its label. Therefore, $BK0$ is a benchmark without DK tasks, $BK25$ has 1 dominant kernel task and 3 dominant transfers tasks (25\%), and so on.


Each experiment has tested all possible tasks permutations for each benchmark (24 different tasks orders are possible). Figure \ref{fig:prediction_error} shows the average prediction error obtained per benchmark and device. It can be observed  that the geometric mean prediction error for all the benchmarks is below 1\% for AMD R9 and NVIDIA K20c devices and 1.12\% for Intel Xeon Phi. This way, we can conclude that our prediction model achieves high accuracy in the calculation of the total execution time of an arbitrary task group ($TG$).

\begin{table}[]
\centering \small
\setlength{\tabcolsep}{0.2em}
\begin{tabular}{|c|c|c|c|c|c|c|c|}
\hline
Device                                                  & \begin{tabular}[c]{@{}c@{}}Comp.\\ Units\end{tabular} & \begin{tabular}[c]{@{}c@{}}DMA \\ Engine\end{tabular} & \begin{tabular}[c]{@{}c@{}}Max Work \\ Group Size\end{tabular} & \begin{tabular}[c]{@{}c@{}}Local Mem.\\ Size (KB)\end{tabular} & \begin{tabular}[c]{@{}c@{}}Global Mem. \\ Size (GB)\end{tabular} & \begin{tabular}[c]{@{}c@{}}PCI\\ Express\end{tabular} & \begin{tabular}[c]{@{}c@{}}OpenCL \\ version\end{tabular} \\ \hline
AMD R9                                                  & 44                                                      & 2                                                     & 256                                                            & 32                                                            & 4                                                                  & 2.0                                                   & 2.0                                                       \\ \hline
\begin{tabular}[c]{@{}c@{}}Xeon Phi 5100\end{tabular} & 236                                                     & 1                                                     & 8192                                                           & $32^*$                                                            & 6                                                                  & 2.0                                                   & 1.2                                                       \\ \hline
NVIDIA K20c & 13                                                     & 2                                                     & 1024                                                           & 48                                                            & 4                                                                 & 2.0                                                   & 1.2                                                       \\ \hline
\end{tabular}
\caption{Overview of the evaluation platforms. $^*$Xeon Phi uses cache and regular GDR memory to simulate local shared memory.}
\label{tb:platforms}
\end{table}

\begin{lstlisting}[language=C++, caption={OpenCL Kernel code for a synthetic task.}, label={lst:SyntheticKernel}]
__kernel void synthetic_kernel(__global int *input, int num_iterations, int factor)
{
int idx = get_global_id(0);

for(int i=0;i<num_iterations;i++)
    input[idx] *= factor;
}
\end{lstlisting}

\begin{table}[ht]
\begin{minipage}[b]{0.56\linewidth}
\centering \small
\setlength{\tabcolsep}{0.2em}
\begin{tabular}{|c|c|c|c|c|c|c|c|c|c|}
\hline
\begin{tabular}[c]{@{}c@{}}Synthetic\\ Task\end{tabular} & T0 & T1 & T2 & T3 & T4 & T5 & T6 & T7 \\ \hline
HTD                                                              & 0.1 & 0.2 & 0.3 & 0.1 & 0.6 & 0.2 & 0.4 & 0.8 \\ \hline
K                                                                & 0.8 & 0.7 & 0.6 & 0.7 & 0.2 & 0.2 & 0.2 & 0.1 \\ \hline
DTH                                                              & 0.1 & 0.1 & 0.1 & 0.2 & 0.2 & 0.6 & 0.4 & 0.1 \\ \hline
\end{tabular}
\caption{Synthetic tasks used in our benchmarks. The {\em HtD}, {\em K} and {\em DtH} commands of each task are defined with a fraction of time with respect to a time unit. The time unit is 10 ms.}
\label{tbl:synthTasks}
\end{minipage}\hfill
\begin{minipage}[b]{0.4\linewidth}
\centering \small
\setlength{\tabcolsep}{0.2em}
\begin{tabular}{|c|c|}
\hline
Benchmark & Tasks              \\ \hline
BK0       & T6, T7, T4, T5 \\ \hline
BK25      & T0, T4, T6, T7 \\ \hline
BK50      & T0, T1, T4, T5 \\ \hline
BK75      & T0, T1, T2, T4 \\ \hline
BK100     & T0, T1, T2, T3 \\ \hline
\end{tabular}
\caption{Synthetic Benchmarks used. Each benchmark is defined with the label BKX, where X is the percentage of dominant kernel tasks in the benchmark.}
\label{tbl:synthBenchmarks}
\end{minipage}
\end{table}

\section{Selecting a tasks ordering}
\label{sec:ExecutionOrder}
In the previous section we showed our model can be used to accurately estimate the execution time of a $TG$. Due to the overlapping of different commands, the execution order can have an important impact on the total execution time of the tasks. In this section we develop a runtime heuristic that is able to establish a near-optimal task ordering that significantly reduces the total execution time of the $TG$.

\subsection{Task reordering heuristic}
\label{sec:heuristic}

The search for an optimum order when offloading a $TG$ onto an accelerator must be performed at runtime. Consequently, methods based on brute force approaches are not applicable. In this section we propose a heuristic which tries to minimizes the inactivity periods in a device when a group of tasks is submitted. The heuristic searches for the highest number of commands which can be concurrently executed according to the device hardware restrictions. In order to achieve this aim, Algorithm ~\ref{alg:Reorder} is proposed.

\begin{algorithm}
\small
RT $= \{T_0,...,T_N\}$: Set of remaining tasks to order.\\
OT $=\{\phi\}$: Set of ordered tasks.\\
t\_HTD$ = 0$: Completion time of the last HTD command.\\
t\_K$ = 0$: Completion time of the last K command.\\
t\_DTH$ = 0$: Completion time of the last DTH command.\\
\caption{Batch Reordering Algorithm}\label{alg:Reorder}
\begin{algorithmic}[1]
\Procedure{Batch Reordering}{}
\State $T_{ini} = select\_first\_task(RT)$
\State $OT = OT \cup \{T_{ini}\}$
\State $RT = RT \: \backslash \: \{T_{ini}\}$
\State $[t\_HTD, t\_K, t\_DTH] = update(OT)$
\While {$|RT| > 2$}
\State $T_{next} = select\_next\_task(RT, t\_HTD, t\_K, t\_DTH)$
\State $OT = OT \cup \{T_{next}\}$
\State $RT = RT \: \backslash \: \{T_{next}\}$
\State $[t\_HTD, t\_K, t\_DTH] = update(OT)$
\EndWhile
\State \textbf{end}
\State $[T_{before \: last}, T_{last}] = select\_last\_tasks(RT, OT)$
\State $OT = OT \cup \{T_{before \: last}, T_{last}\}$
\\
\textbf{end}
\EndProcedure
\end{algorithmic}

\end{algorithm}

The heuristic starts with the selection of the first task (Algorithm ~\ref{alg:Reorder} line: 2). It is selected among tasks with a short {\em HtD} command and a long {\em K} command when compared to the remaining tasks in $RT$ (set of remaining task to order).  This way, GPU inactivity is reduced at the beginning of the execution and, in addition, overlapping options for the following tasks are leveraged. If several tasks fulfill these requirements, then the task with longer {\em DtH} command is selected so that the concurrency among the transfer and kernel commands is improved. The first selected task, $T_{ini}$, is added to $OT$ (set ot ordered tasks) and removed from $RT$ (lines: 3-4). Then, $t\_HTD$, $t\_K$ and $t\_DTH$ can be updated by simulating the ordered tasks set $OT$ (line: 5).

While the number of tasks in the $RT$ set is higher than 2, subsequent tasks are selected (lines: 6-11). This selection is accomplished by looking for the best fit between the remaining {\em K} commands of the previous selected tasks and the {\em HtD} command of the new task, and between the remaining {\em DtH} commands of the previously selected tasks and the {\em K} command of the new task (line: 7). More precisely, our execution model is used to predict the computation time of the current commands in $OT$, to compare them with the execution time of the new tasks, and to maximize the overlapping degree among the commands.

When the number of tasks in $RT$ set is equal to 2, the selection of the last task is carried out by the $select\_last\_tasks$ function (line: 12). This function works as $select\_next\_function$ but adding a new criterion based on the duration of the {\em DtH} command, to avoid a long accelerator inactivity period during the execution of the {\em DtH} command of the last task.

Note that global memory requirements for concurrent scheduling of a $TG$ can be higher than those when the tasks are sequentially executed, as several running tasks can need to simultaneously allocate global memory for both input and output data. Consequently, the maximum storage allocation can be an additional restriction when selecting the tasks forming part of a $TG$. An efficient and flexible solution for this problem may need the development of a specific global memory allocation policy on the device. This aspect is not studied in this paper because it is focused on testing the heuristic performance when many independent tasks are available. Consequently, we  assume that there is enough available global memory to allocate input and output data required by the concurrent tasks.

\section{Experimental results}
\label{sec:results}

In this section our tasks execution model and heuristic are evaluated. For that, some benchmarks composed of real tasks are tested along with the synthetic benchmarks presented in Subsection~\ref{sec:model_val}. Then, a multi-threaded scenario where several threads running on the host launch tasks on the accelerator is evaluated.

\subsection{Real tasks benchmarks}

The real tasks benchmarks have been built with several well-known kernels belonging to NVIDIA and AMD OpenCL SDK. Tables~\ref{tbl:real_tasks} summarizes the selected kernels alongside its classification as dominant kernel or transfer.  As it can be observed in the table, DCT and FWT tasks can present a different behaviour according to the used device. In order to increase the variability of the benchmarks, each task has been executed using several data sizes leading to different execution times. In Table~\ref{tbl:real_tasks_commands} the ranges of execution times for the commands belonging to the real tasks using different data sizes are shown.

The real tasks have been combined in several benchmarks similarly to what it was done with synthetic tasks in Section~\ref{sec:model_val}, but using the real tasks from Table~\ref{tbl:real_tasks}. Therefore, in benchmark $BK0$ every task is dominant transfer, in $BK25$ a 25\% of the tasks have dominant kernels, in benchmark $BK50$ there are dominant kernel and dominant transfers tasks equally distributed, in benchmark $BK75$ a 75\% of the tasks have dominant kernels, and in benchmark $BK100$ every task is dominant kernel.

\begin{table}[htpb]
\centering \small
\setlength{\tabcolsep}{0.15em}
\begin{tabular}{|c|c|c|}
\hline
Kernel & Description           & Type of Task \\ \hline
MM     & Matrix Multiplication & DK           \\ \hline
BS     & Black Scholes         & DK           \\ \hline
FWT    & Fast Walsh Transform  & DT/DK           \\ \hline
FLW    & Floyd Warshall        & DK           \\ \hline
CONV   & Separable Convolution & DK           \\ \hline
VA     & Vector Addition       & DT           \\ \hline
TM     & Matrix Transposition  & DT           \\ \hline
DCT    & Discrete cosine transform  & DT/DK           \\ \hline
\end{tabular}
\caption{Tasks used in the real benchmarks. The tasks have been selected according to its classification as dominant kernel or transfer. DCT and FWT tasks can have a different behaviour according to the device. Thus, both tasks can be dominant transfer or dominant kernel, if they are executed on AMD R9 and NVIDIA K20, or Intel Xeon Phi respectively.}
\label{tbl:real_tasks}
\end{table}

\begin{table}[t]
\centering \tiny
\setlength{\tabcolsep}{0.2em}
\begin{tabular}{|c||c|c|c|c|c|c|c|c|c|}
\hline
\multirow{2}{*}{\textbf{Device}}  &
\textbf{Task}                                                  & \textbf{MM}        & \textbf{BS}        & \textbf{FWT}       & \textbf{FLW}       & \textbf{CONV}       & \textbf{VA}        & \textbf{MT}       & \textbf{DCT}       \\ \cline{2-10}
& Dominance                                                                        & DK        & DK         & DK/DT        & DK         & DK         & DT        & DT       & DK/DT     \\ \hline

\multirow{3}{*}{\begin{tabular}[c]{@{}c@{}}AMD \\ R9\end{tabular}} &
\begin{tabular}[c]{@{}c@{}}HtD\\ (ms)\end{tabular}    & 0.97-2.57 & 0.08-1.29 & 1.29-2.57 & 0.05-0.07  & 0.09-0.37  & 0.65-3.86 & 2.57-5.15 & 2.57-5.15 \\ \cline{2-10}
& \begin{tabular}[c]{@{}c@{}}Kernel\\ (ms)\end{tabular} & 1.80-9.02 & 2.98-5.57 & 2.59-5.47 & 7.77-10.08 & 1.51-14.58 & 0.05-0.30 & 0.29-3.59 & 0.95-1.89 \\ \cline{2-10}
& \begin{tabular}[c]{@{}c@{}}DtH\\ (ms)\end{tabular}    & 0.14-1.18 & 0.16-2.17 & 1.18-2.35 & 0.09-0.16  & 0.09-0.37  & 0.30-1.81 & 2.36-4.70 & 2.35-4.71  \\ \hline
\multirow {3}{*}{\begin{tabular}[c]{@{}c@{}}Intel Xeon \\ Phi\end{tabular}} &
\begin{tabular}[c]{@{}c@{}}HtD\\ (ms)\end{tabular}    & 0.36-0.90 & 0.17-0.63
& 0.67-1.26 & 0.03-0.06  & 0.06-0.17 & 1.27-7.46 & 2.58-4.98 & 1.71-2.25\\ \cline{2-10}
&\begin{tabular}[c]{@{}c@{}}Kernel\\ (ms)\end{tabular} & 4.98-5.03 & 5.25-12.03 & 4.59-6.39 & 1.12-9.05 & 0.56-10.09 & 0.18-1.18 & 2.36-1.09 & 6.97-9.41\\ \cline{2-10}
& \begin{tabular}[c]{@{}c@{}}DtH\\ (ms)\end{tabular}    & 0.09-0.16 & 0.33-1.24 & 0.61-1.21 & 0.06-0.12 & 0.17-10.09 & 0.61-3.68 & 2.54-4.93 & 1.67-2.18\\ \hline
\multirow {3}{*}{\begin{tabular}[c]{@{}c@{}}NVIDIA \\ K20c\end{tabular}} &
\begin{tabular}[c]{@{}c@{}}HtD\\ (ms)\end{tabular}    & 2.51-3.77 & 0.31-1.25
& 1.25-5.01 & 0.01-0.31  & 0.63-2.53 & 2.51-12.54 & 2.60-5.01 & 2.51-5.01\\ \cline{2-10}
&\begin{tabular}[c]{@{}c@{}}Kernel\\ (ms)\end{tabular} & 3.99-7.95 & 1.25-9.26 & 1.20-4.94 & 1.32-9.25 & 1.47-9.20 & 0.09-0.44 & 0.41-2.61 & 1.55-3.08\\ \cline{2-10}
& \begin{tabular}[c]{@{}c@{}}DtH\\ (ms)\end{tabular}    & 1.24-2.49 & 0.62-2.50 & 1.25-4.98 & 0.03-0.63 & 0.62-2.50 & 1.25-6.19 & 2.60-4.96 & 2.48-4.96\\ \hline
\end{tabular}
\caption{Range of execution times for {\em HtD}, {\em K} and {\em DtH} commands per each real task. The range of execution times for each command from the tasks is obtained executing the corresponding task with different data size.
}
\label{tbl:real_tasks_commands}
\end{table}

\subsection{Execution of many tasks by several threads}

\begin{figure}[ht]
  \centering
  \includegraphics[width=0.9\linewidth, height=0.2\linewidth]{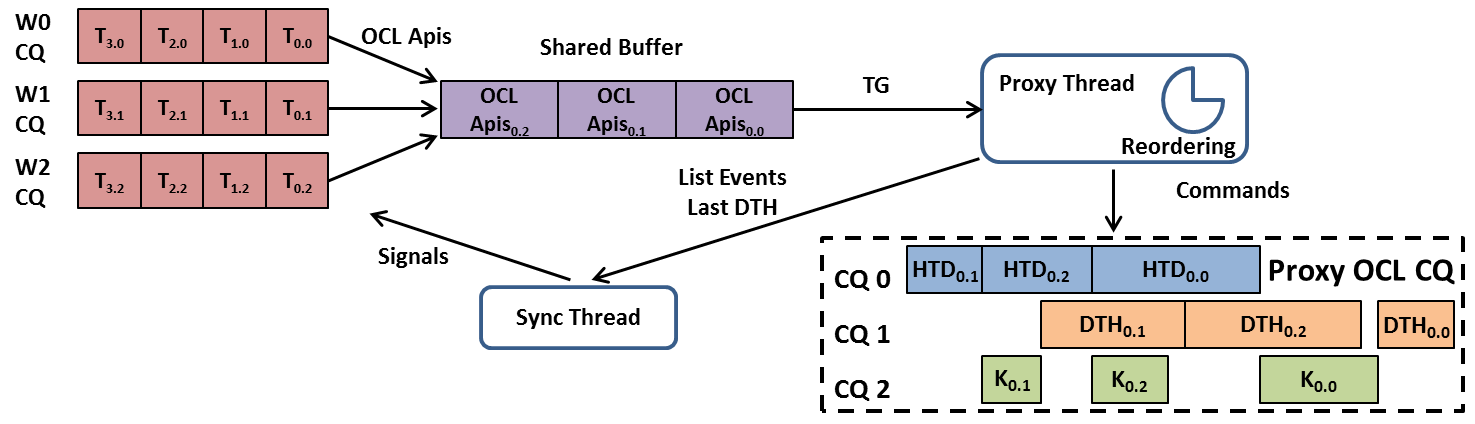}
  \caption{The proxy thread is in charge of reordering the $TG$ found in the buffer and submitting the corresponding commands to the CQs. The figure shows three threads submitting OpenCL APIs to different command queues (CQ). These APIs are intercepted and written in a shared buffer. Proxy thread samples this buffer, establishes the $TG$ and reorders the commands according to the proposed heuristic. Finally the reordered commands are submitted by the proxy thread to the accelerator employing three CQs (a two DMA engines device is assumed in this example).}.
  \label{fig:Proxy Thread}
\end{figure}

Our tasks execution model and heuristic have been evaluated in a demanding multithreaded scenario where several threads running applications (workers) offload one or several tasks onto a device. All workers send task information regarding OpenCL API calls to a buffer that is constantly polled by a host proxy thread. As it can be seen in Figure \ref{fig:Proxy Thread}, this thread is in charge of reordering the $TG$ found in the buffer and submitting the corresponding commands to the CQs. In the experiments we have conducted, the workers are also running on the host. This way the communication between workers and the host proxy thread can be easily implemented through shared memory. This simple scenario allows us to focus on the main subject of this paper, that is, the impact of task reordering in the execution of concurrent tasks.


In the experiments we consider $T$ sets of independent tasks with $T$ taking values of 4, 6 and 8. In each set, a batch of $N$ dependent tasks is available with $N$ taking values of 1, 2 or 4. The $T \cdot N$ tasks are randomly selected from the corresponding synthetic or real benchmark. Two experimental setups are defined to establish how good is the order calculated by the heuristic compared to all possible tasks orders. These configurations are named Heuristic Setup and NoReorder Setup:

\begin{figure}[ht]
\centering
\includegraphics[width=130mm]{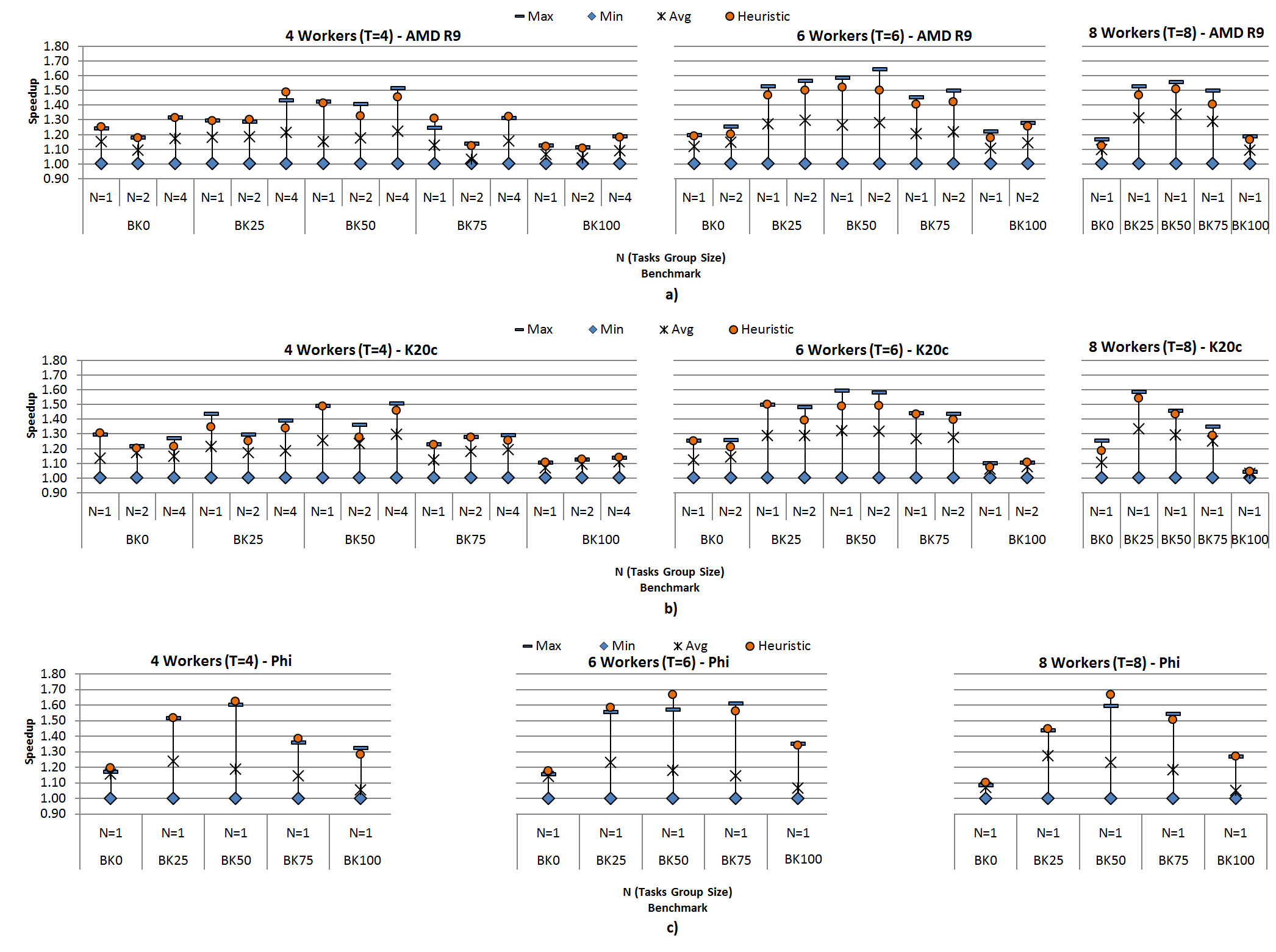}
\caption{Speedups achieved with synthetic tasks, for every benchmark in AMD R9 (at the top), NVIDIA K20c (in the middle) and Intel Xeon Phi (bottom) devices, with respect to the worst permutation. Maximum speed-up is achieved by the best permutation, median speed-up is obtained using the median execution time, and heuristic speed-up is computed using the ordering obtained by our heuristic.} \label{fig:syntheticTasks}
\end{figure}

{\bf NoReorder setup}. A thread asynchronously submits the commands of $T \cdot N$ tasks taking into account the imposed dependencies between task belonging to the same batch. Unlike \cite{Lazaro17}, where only a CQ is employed to submit kernel commands, our approach enables CKE by using a CQ per kernel command. Then, each experiment randomly selects the $T \cdot N$ tasks and carries out fifteen executions of all possible tasks permutations ($(T!)^N$). Notice that given a specific permutation no reordering is applied to the tasks forming part of that permutation. The median execution time per permutation is extracted.

{\bf Heuristic setup}. This setup considers $T$ worker threads launching $N$ consecutive tasks per thread. Thus, the maximum number of concurrent tasks in a $TG$ is $T$. For each experiment, the same tasks selected for the NoReorder setup are employed. Workers write OpenCL API calls for launching the tasks in a common buffer. Dependencies between the tasks launched by a worker are enforced by imposing that a new task is not written in the buffer until the previous task has completely finished. The host proxy thread reads the common buffer and applies the heuristic to calculate a better tasks order.
Next, the commands of the reordered tasks are submitted. Finally, once the host proxy thread submits the $HtD$ command of the last task belonging to the current $TG$, it polls again the common buffer and repeats the cycle. Fifteen executions of the reordered tasks are performed and the median value is extracted.
An example of this setup with $T$=3 and $N$=4 is shown in Figure~\ref{fig:Proxy Thread}.

\begin{figure}[ht]
\centering
\includegraphics[width=130mm]{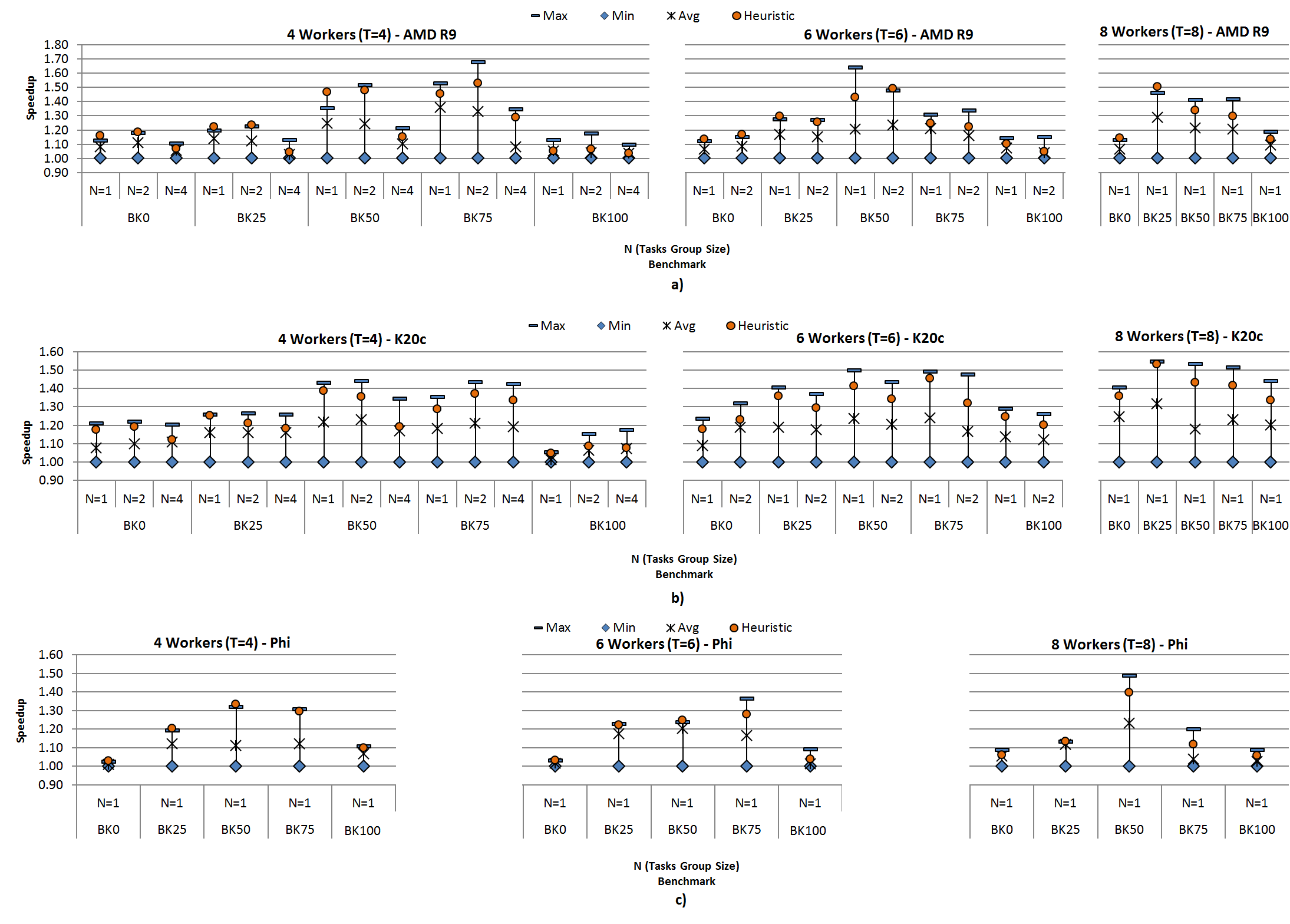}
\caption{Speedups achieved with real tasks, for every benchmark in AMD R9 (at the top), NVIDIA K20c (in the middle) and Intel Xeon Phi (bottom) devices, with respect to the worst permutation. Maximum speed-up is achieved by the best permutation, median speed-up is obtained using the median execution time, and heuristic speed-up is computed using the ordering obtained by our heuristic.} \label{fig:realTasks}
\end{figure}

Figure~\ref{fig:syntheticTasks}  and  Figure~\ref{fig:realTasks} depict the achieved results by synthetic and real benchmarks in AMD R9, NVIDIA K20c and  Intel Xeon Phi devices respectively. The results show the speedup achieved by the geometric mean (cross symbol) and the minimum (blue rectangle) execution times of the NoReorder setup with respect to the maximum execution time (blue rhombus with speedup equal to one) of the same setup. This way, we can visualize the range of speedup values achieved for all possibles task orders permutations in the NoReorder setup (vertical segment with blue rectangle and rhombus end points). All possible permutations have been evaluated for the NoReorder setup using four workers ($T$=4) and $N$=1, 2 and 4. In case of $T$=6, all the permutations are run for $N$=1 but only a subset, randomly chosen, containing 5\% of all possible permutations are used for $N$=2. For $T$=8, $N$ only takes a value of 1 due to the large number of available permutations for $N$ higher than 1. As  Xeon Phi has only one DMA engine, experiments have been conducted with all the possible permutations for $N$=1 ($N$=2 and $N$=4 produce the same speedup results). In addition, the speedup achieved by the order calculated by our heuristic (red circle) with respect to the experiment giving the maximum execution time in the NoReorder setup is indicated.

\begin{figure}[t]
  \centering
  \includegraphics[width=0.7\linewidth]{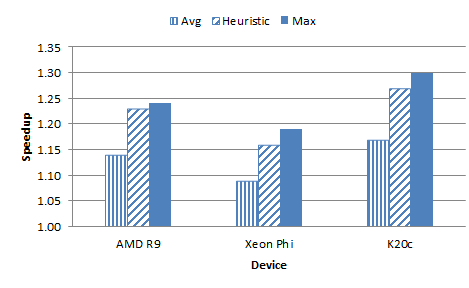}
  \caption{Geometric mean of the Maximum, Average and Heuristic speedup for the real tasks experiments.}
  \label{fig:MeanSpeedups}
\end{figure}

\begin{table}[]
\centering
\begin{tabular}{|c|c|c|c|}
\hline
$T$ (Number of concurrent tasks)                                             & 4 & 6 & 8 \\ \hline
\begin{tabular}[c]{@{}c@{}}Avg. CPU Scheduling Time (ms)\end{tabular} & 0.06  & 0.10  & 0.22  \\ \hline
\begin{tabular}[c]{@{}c@{}}Avg. Device Execution Time (ms)\end{tabular}  & 28.04 & 37.82 & 49.78 \\ \hline
\end{tabular}
\caption{Average scheduling overhead incurred by the host proxy thread running on an Intel Core 2 Quad for benchmarks with 4, 6 and 8 synthetic tasks. The average execution time of a group of 4, 6 and 8 tasks on a K20c is also shown.}
\label{tbl:proxy-overhead}
\end{table}

For synthetic benchmarks, Figures~\ref{fig:syntheticTasks}.a (AMD R9), \ref{fig:syntheticTasks}.b (NVIDIA K20c) and \ref{fig:syntheticTasks}.c (Intel Xeon Phi) show that, as expected, the impact of tasks reordering is higher for BK25, BK50, BK75 benchmarks as higher speedups are achieved. The reason behind these behaviour is that these benchmarks contain different types of tasks (transfer and kernel dominant) ergo better opportunities for command overlapping can be found out. In addition it can be observed that our heuristic predicts orderings very close to the best permutation most of the time for any benchmark and $T$ and $N$ values, and always better than the mean execution time achieved by the NoReorder setup. There are cases in AMD R9 (e.g. $BK25$, $T=4$, $N=4$) and Xeon Phi (e.g. $BK50$, $T=6$, $N=1$) where our heuristic is able to perform better than the best permutation of the NoReorder setup. After a study of this specific results we have concluded that the use of $CKE$ sometimes hinders the kernels execution of the $TG$, increasing its total execution time w.r.t a non-CKE configuration. Additionally, we have also evaluated the overhead incurred by the heuristic when different values of $T$ (number of concurrent tasks) are employed. Table \ref{tbl:proxy-overhead} shows the time spent by the heuristic (average CPU scheduling time) and the average time taken by the execution of the concurrent tasks after applying the heuristic (average device execution time). It can be seen that the overhead is always below 0.4\%.

Similar conclusions can be extracted from the real benchmarks results shown in Figure~\ref{fig:realTasks}: our heuristic is able to improve the mean value obtained by NoReorder setup and many times it is able to reach a speedup very close to the best value accomplished by NoReorder. There are a few cases where although the heuristic speedup value is better than the average, it is far from the best value obtained by NoReorder setup (e.g. $BK50$ with $T=4$ and $N=4$ for K20c device). In these cases HyperQ is able to find overlapping opportunities among commands belonging to different $TG$ while our heuristic only works with commands belonging to a specific $TG$. Despite that, the average speedup achieved by the heuristic for all the cases is still very high as it is shown in Figure \ref{fig:MeanSpeedups}. Thus, for AMD R9 our heuristic obtains an average speedup of 1.23 which is a 96\% of the improvement obtained by the best ordering of the NoReorder setup (1.24). Finally, the marks obtained in Xeon Phi, 1.16 (84\%),  and K20c, 1.27 (87\%) indicates that our heuristic is able to find a near-optimal ordering.



\section{Conclusions}
\label{sec:conclusion}
We have presented a new strategy that takes advantage of OpenCL support for concurrent command execution on accelerators to increase tasks throughput. This study is motivated by the fact that, given a set of concurrent independent tasks to be executed in an accelerator, the tasks offloading order on the device can have an important impact on the total execution time of the tasks. Our approach proposes a runtime heuristic based on a task execution model that has also been developed in this paper. The heuristic has been successfully tested within an exhaustive setup. This setup generates all the possible task orderings (permutations) allowing not only the overlapping among the transfers and kernel commands but also CKE. The experimental results have been conducted on three different accelerator devices (AMD R9, NVIDIA K20c and Intel Xeon Phi) employing real benchmarks in order to analyze the applicability and generality of the proposed approach. Experiments indicate the heuristic is able to find always an ordering with a better execution time than the average of the permutations and, most times, it achieves a near-optimal ordering (very close to the execution time of the best permutation) with a negligible overhead.

In future works we plan to extend our tasks execution model to include kernels that can be concurrently executed in the accelerator, in order to obtain more overlapping opportunities. We would also like to integrate our heuristic and execution model in a multi-GPU architecture to improve tasks scheduling in this type of systems.

%
\label{sect:bib}
\bibliographystyle{plain}
\bibliography{acmart}


\end{document}